*Article*

# Virtual Astronaut for Scientific Visualization—A Prototype for Santa Maria Crater on Mars

**Jue Wang \*, Keith J. Bennett and Edward A. Guinness**

Department of Earth and Planetary Sciences, Washington University in St. Louis, 1 Brookings Drive, CB 1169, St. Louis, MO 63130, USA; E-Mails: bennett@wunder.wustl.edu (K.J.B.); guinness@wunder.wustl.edu (E.A.G.)

**\*** Author to whom correspondence should be addressed; E-Mail: wang@wunder.wustl.edu; Tel.: +1-314-935-7033; Fax: +1-314-935-4998.

**Abstract:** To support scientific visualization of multiple-mission data from Mars, the Virtual Astronaut (VA) creates an interactive virtual 3D environment built on the Unity3D Game Engine. A prototype study was conducted based on orbital and Opportunity Rover data covering Santa Maria Crater in Meridiani Planum on Mars. The VA at Santa Maria provides dynamic visual representations of the imaging, compositional, and mineralogical information. The VA lets one navigate through the scene and provides geomorphic and geologic contexts for the rover operations. User interactions include *in-situ* observations visualization, feature measurement, and an animation control of rover drives. This paper covers our approach and implementation of the VA system. A brief summary of the prototype system functions and user feedback is also covered. Based on external review and comments by the science community, the prototype at Santa Maria has proven the VA to be an effective tool for virtual geovisual analysis.

## 1. Introduction

There have been a number of successful orbital and landed missions to Mars over the past a few decades. The orbital missions include Mariner 9, Viking orbiter, Mars Global Surveyor (MGS), Mars Odyssey, European Space Agency (ESA)'s Mars Express (MEX), and Mars Reconnaissance Orbiter (MRO) missions. The landed missions include Viking, Pathfinder, Mars Exploration Rover (MER), Phoenix, and the Mars Science Laboratory (MSL) missions. Terabytes of imaging and other data have been acquired from those Mars missions. The size of the image database keeps growing with the current ongoing Odyssey, MRO, MEX, MER and MSL missions. More images are expected from future missions. With the huge amount of high-quality Mars data coming to the planetary science community and the general public, how to help people, especially the geologists, utilize the data has become an important topic.

Several interactive tools or online systems have been developed at different institutions to search and visualize the Mars data in 2D [1–5]. The 2D visualization adds the geographic, visual dimension to the data, but lacks the depth of a 3D visualization. The 3D Visualization of topographic models with commercial tools, such as ArcScene from ESRI® ArcGIS, ENVI, ERDAS Imagine, and Surfer provides a better sense of the Mars terrain. These tools let users manipulate 3D models with operations like translation, rotation, zoom in or zoom out. These bring out insights into the data that are not possible with 2D visualizations. For example, observing the subtle changes in elevations with a 3D model. However, the software based 3D visualization on a desktop has no immersion capability and limited interactive functions. The application of virtual reality (VR) techniques to planetary data sets started in early 1985 at NASA AMES Research Center. The Virtual Planetary Exploration (VPE) lab of AMES built a head-mounted, wide-angle, stereoscopic display system to interactively explore the surface of Mars based on Viking images [6,7]. Building a VR system with navigation, interaction, and/or immersion capabilities for scientific visualization for Mars is cost-effective relative to the high cost and the unfriendly Mars environment of a manned mission to Mars. Using VR techniques to build a computer-simulated 3D environment with remote sensing data at various resolutions will help the study of the environment and evolution of this planet.

In the early 1990s, the concept of a Cave Automatic Virtual Environment (CAVE) was developed at the University of Illinois at Chicago's Electronic Visualization Laboratory. CAVE provides an immersive environment by projecting images to multiple walls of a room-sized cube. Various CAVEs have been developed in manufacturing, architecture, urban planning, medicine, simulation, and education. ADVISER (Advanced Visualization in Solar System Exploration and Research) from Brown University was a CAVE application developed for immersive scientific visualization applied to Mars research and exploration [8]. The Fossett Laboratory in the Department of Earth and Planetary science at Washington University in St. Louis also developed a CAVE system to support the planetary studies, including decision making for the MER and Phoenix missions. However, considering the size and cost of a CAVE system, a desktop VR system is more affordable to the researchers and geologists.

Desktop VRs have been developed for various planetary applications. Oravetz *et al.* [9] conducted a study about the perceptual errors humans made while judging lunar-like terrain or lunar terrain built from test data and Apollo mission images. They concluded the need to develop a prior-mission VR training tool to calibrate an astronaut's slope, distance, and height perception that would help overcome future navigational difficulties. A virtual microscope library was developed for online delivery of extraterrestrial samples such as lunar samples and Martian meteorites. This project was mainly developed for public engagement and outreach purpose to stimulate the public's interest in planetary and space sciences [10]. The virtual "Thinking Journey to Mars" program was developed using the MGS and the Mars Pathfinder data in a 3D VR system to educate the high-school students in Israel. The program helped students to visualize the abstract and alien landscapes with images, animations and simulations [11]. Other educational virtual environments to support the teaching of planetary phenomena were introduced in [12,13].

In addition, several computer-based 3D VR simulators were developed for testing, validation and operation of planetary rovers. A detailed summary of the development of robotic exploration systems was given in [14]. VEVI (Virtual Environment Vehicle Interface, [15]) was the earliest robotic operational system developed since 1990s. Viz [14] was a visualization tool designed for the Mars Polar Lander mission based on previous experience with VEVI, MarsMap (for Mars Pathfinder mission, [16]), C-Map (the Chernobyl Mapping System, [17]), and VDI (Virtual Dashboard Interface, [18]). Viz was further customized for VIPER (Virtual Intelligent Planetary Exploration Rover, [19]). A 3D VROS (Virtual Rover Operation Simulator, [20]) and RSVE (Rover Simulation Environment [21]) were also developed recently by research groups in China.

A Virtual Astronaut (VA) has been developed for scientific exploration at the Geosciences Node of NASA's Planetary Data System (PDS). Unlike the systems for the rover simulation, education, or outreach purpose as introduced above, the major objective of the VA is to develop a precise, highly-detailed, scientifically accurate 3D virtual reality environment that allows users to not only view observations made by a rover from the rover's perspective, but also allows them to view the scene and data from any perspective they want. In effect, the VA allows scientists to view a location of another planet in a manner similar to the way geologists would investigate the area if they were actually present there. The VA tool also combines images and topographic data obtained from orbital spacecraft with the rover-based observations to provide a regional context to the area being investigated by the rover. The VA is targeted at the planetary science community and is designed to run as a desktop application or within a web browser.

An initial version of the VA was built for the MER Opportunity's campaign at Santa Maria Crater on Mars. Figure 1 presents a virtual view of the VA with multiple image mosaics overlaid on a digital elevation model. The functionality and features of the VA evolved during design and development based on testing and feedback from many potential science users. Other versions of the VA are planned for Opportunity's exploration of Endeavour Crater, including several areas of its rim along Cape York, and selected areas studied by the Mars Science Laboratory rover. In this paper we describe the general approach and challenges for building the VA including how user feedback helped refine its functionality. Then, the specific aspects of the Santa Maria version are detailed.

**Figure 1.** View of the Virtual Astronaut is shown. Icons in the upper left are for selecting the tools' functions. Santa Maria Crater is located at the center of the view. The image data is a combination of orbital HiRISE (High Resolution Imaging Science Experiment) data from Mars Reconnaissance Orbiter (MRO) mission and ground-based Panoramic Camera (Pancam) and Navigation Camera (Navcam) images acquired by Opportunity. The Opportunity Rover is located in the red circle in the figure.

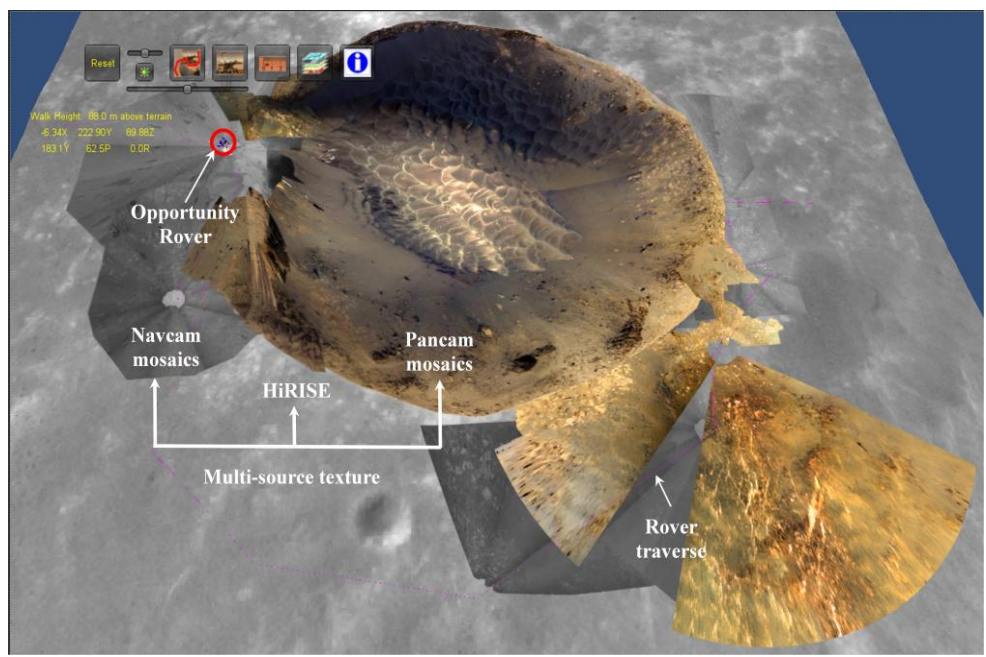

## 2. Related Work

A few desktop virtual globes are freely downloadable for scientists and the general public for Mars exploration, including Mars in Google Earth [22], Microsoft's WorldWide Telescope (WWT) [23], and NASA World Wind from NASA AMES Research Center [24]. Blaschke *et al.* [25] gave a review on the technical developments and standards of the virtual globes. These tools give excellent global views of Mars from orbit. They also provide the capability to visit a specific region by clicking any point in the globe.

However, the user might not identify detailed features at centimeter levels when one goes to a small local region, because of the relative lower resolution base maps being applied in these tools. NASA World Wind displays MGS Mars Orbiter Camera (MOC) Narrow Angle (NA) images with a resolution from 1.5 to 12 m/pixel on the globe. The best available orbital data with Mars in Google Earth and WWT are HiRISE images, which can see down to about 0.25 m per pixel [26]. Both MOC NA and HiRISE data lack a complete coverage on Mars. The pixelated texture or a blank scene requires adding other data such as ground images into the virtual globes. None of the tools seems to integrate the ground data with the orbital data to generate a high-resolution basemap. Some of the tools just provide options to pop-up image mosaics taken by the MER rovers. Others provide functions for users to place customized 2D Web Mapping Service (WMS) image overlays, but the task is non-trivial for general users without programming, cartography, GIS, and image processing background.

A number of projects have come up with the VR applications for small areas of Mars. These VRs can be used interactively via the web with network improvements. Exploration guides to Gale Crater, Spirit's Journey, and Curiosity's Journey [27] were developed as part of NASA/JPL/CalTech's outreach program. These tools are good introductions to Mars and the rover missions for the public. For example, Spirit's Journey tells the user what the Spirit Rover saw using pop-ups of rover images, how it drove, and how far it went. However, these tools lack high-resolution surface textures and detailed data and information needed for more-complete scientific studies.

The above-mentioned VR systems are high-quality visualization tools for planetary image data sets. These tools tend to either focus on orbiter or ground-based data, but do not combine both types of data in the surface texture. This research will build a virtual environment by integrating the real orbital and landed mission data to provide users the context from various perspectives, including a rover's.

## 3. Building the Virtual Astronaut

The VA is built on the Unity3D Game Engine and development environment (Version 3.4). Figure 2 shows the process for building the VA for a region of interest.

**Figure 2.** The steps for building a Virtual Astronaut for a specific region are depicted. Pre-processing combines and registers multiple data sets into a uniform format. The interactive tool is based on custom software working with standard Unity modules.

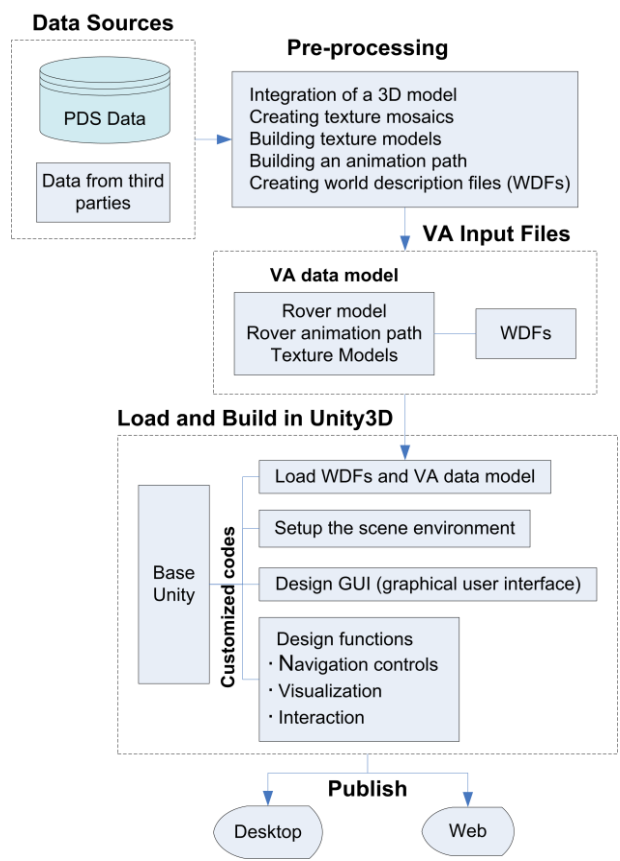

Multiple orbital and surface data sets obtained from the NASA's PDS or third parties are pre-processed to build a VA data model using a combination of tools. The VA data model includes a texture model, which is a 3D model of the surface with texture derived from image mosaics, animation paths, and a rover model. A set of world description files (WDFs) are created to describe the objects of a VA data model. This data pre-processing is the key step of generating a specific instance of the VA. Once generated, the VA data model and WDFs are loaded into a Unity3D-based virtual environment. A set of in-house tools and procedures have been developed for visualization and interactive functions within the VA, supplementing the basic modules and functions provided by Unity. These include visualization, navigation and interactive functions that evolved based on user feedback. The final application is published as a desktop application and on the web.

## 3.1. Data Pre-Processing and Challenges

Data pre-processing to put the multiple data sets into a common base is necessary for building a VA data model. The VA data model includes a texture model, animation paths, and a rover model. The elements of the data model are tied together by a set of WDFs stored as comma-separated value (CSV) files.

The texture model includes a 3D terrain model with several image mosaics mapped onto the terrain to provide surface texture. The large amount of rover-based image data due to its high-resolution necessitated texture mapping of the images during the pre-processing step because projection on the fly required a large amount of computation power, which affected the overall system performance. Rover-based images are mapped to the terrain by a ray tracing technique. Ray tracing first determined the viewing ray direction of each pixel in the 2D images using camera model parameters, and then calculated the intersection of the viewing ray with the terrain. Mosaicking the textural information from multiple sources and overlaying the texture mosaic on a 3D terrain model were done using ESRI$^{®}$ ArcGIS. The texture model was then exported to a VRML (Virtual Reality Modeling Language [28]) file using ArcScene and was further transformed to a FBX (Autodesk FilmBox format, [29]) file by PolyTrans [30]. The FBX file was brought into the VA environment with customized code interfaced to Unity.

The challenge for generating the texture model is integrating data from multiple sources taken at different resolutions and under different lighting and viewing conditions, and registered to different reference frames. Scale, shape, and location differences between the same features are caused by working with multiple sensors acquiring data under different conditions. Another challenge is registering rover position to features in the texture model. Rover positions are derived from an onboard inertial measurement unit (IMU) and from tracking the number of wheel turns. Rover position and orientation have uncertainties accumulated over time due to the wheel slippage and IMU drift. The rover attitude and position are periodically refined by the measurement of the Sun position [31] or by triangulation with surface features [32–34]. Various coordinate systems are associated with data from orbit and the surface. This requires a clear understanding of the sensor models and manual registration of multiple data sets into the same reference frame. In addition, Unity uses a left-handed Y up coordinate system. This is different from the GIS world where X/Y is the surface plane and Z is usually up. Finally, combining images from various cameras with different lighting conditions often involves manual operations of color adjustment as discussed in Section 4.3.

*3.2. User Inputs and Feedback*

The survey of user requirements is an important factor designing and building the VA. In the initial design stage, lots of discussions were carried out with the students and staff working at the Department of Earth and Planetary Sciences at Washington University in St. Louis. These potential users gave feedback about the most useful functions they would like to see in the VA. For example, a user might like to drive the rover along its traverse or move independently around the virtual surface. In another case, a user might want to examine a place where the Opportunity Rover collected a series of Microscopic Imager (MI) images, or took an Alpha Particle X-ray Spectrometer (APXS) measurement. These requirements lead to the development of interactive functions such as animation control of rover drives, visualizations of *in-situ* observations, and measurement of the size of features. Users described what activities they would like to control during a rover animation. For example, they would like to view the rover model moving along the path taken by Opportunity. Control of the rover model during the animation includes changing its speed or pausing the rover so that the user can take a close look at the area around where the rover is located. These inputs have been helpful in designing the interactive functions for the VA.

There were considerable discussions with potential users between using scene controls *vs.* viewpoint controls for navigation through the scene. Students and geologists familiar with software such as ArcScene or ENVI initially preferred the scene controls. These controls let users manipulate an object, e.g., making a 3D terrain surface rotate around the center and viewed with different observation angles. This was not a realistic experience for a virtual user to walk around the surface. Therefore, after several iterations of user testing, viewpoint controls were selected to achieve more realistic navigation through the 3D environment. The viewpoint controls move the viewpoint forward/backward, up/down, left/right, or turn it left/right, up/down. A user can control the location of the viewpoint as well as its orientation. These controls provide great flexibility to move around and look around the scene. The viewpoint is manipulated by interpreting the screen coordinates of the mouse with respect to the screen center. For example, when one clicks the mouse on the scene, the screen's movement follows the user's mouse as if pointing in the direction of desired movement.

Function control based on menus or icons was also the subject of intensive user testing. A menu-based approach was dropped because the icon-based approach was shown to be easier to operate. The VA had several functions that were prototyped and evaluated with user testing and then not put into the final version due to the user confusion or complexity of a function. We found it was easy and fast to implement functions in the Unity environment. The input devices for navigation were limited to a keyboard, mouse, and gamepad based on an early requirement on the system.

Before release, the VA was tested on a set of machines with various processors, amount of memory, and graphics cards. Feedback such as a frozen screen, the lagging response from operations, and program crashes required reducing of the size of the VA data model, which resulted in a faster loading speed and better performance. The navigation method was revised according to the user experience. The VA originally included two different modes for navigation: fly mode and walk mode. Fly mode allows for a full range of movement, while walk mode restricts motion to a chosen altitude, as if you were walking on Mars. Fly mode was eliminated from the VA for the first release based on user feedback.

The VA was further reviewed by two external reviewers and was demonstrated before a dozen of scientists and specialists during conferences or meetings. The science community gave positive feedback and a few detailed suggestions. More information about sol, site, and drive number as the rover drives over each observation was displayed on the screen upon the user request. We also added the functionality to display full resolution images on demand. A few comments were taken under consideration to improve documentation, control functions and user interface. The VA is continuously being improved upon the user comments and feedback.

## 4. The Virtual Astronaut for Santa Maria

The first release of the Virtual Astronaut covers Opportunity's Santa Maria campaign. Santa Maria is an impact crater on Mars that is about 90m in diameter and is located at 2.172° S, 5.445° W in Meridiani Planum. The crater sits northwest of the much larger Endeavour Crater. Before resuming its long-term trek toward Endeavour, Opportunity investigated Santa Maria from 16 December 2010 to 22 March 2011 (Martian days or sols 2451–2545). During the campaign, Opportunity conducted a number of experiments around Santa Maria. The interior and exterior of the crater were extensively imaged from several viewpoints to study the structure of the crater. These image sequences provided multispectral data. Topography was also derived from the stereo observations. In addition, two rock targets, Luis De Torres and Ruiz Garcia, where investigated in detail with the instruments on the rover's arm. Data collection for the rock targets included close-up images and elemental oxide abundances.

Data integration is an important step to build texture models for the VA. Figure 3 illustrates the data source used and their integration for the VA at Santa Maria Crater. Raw data were originally obtained from the PDS. Derived products were developed in-house and by third parties. These derived products were registered and integrated into a texture model.

**Figure 3.** Data source and data integration.

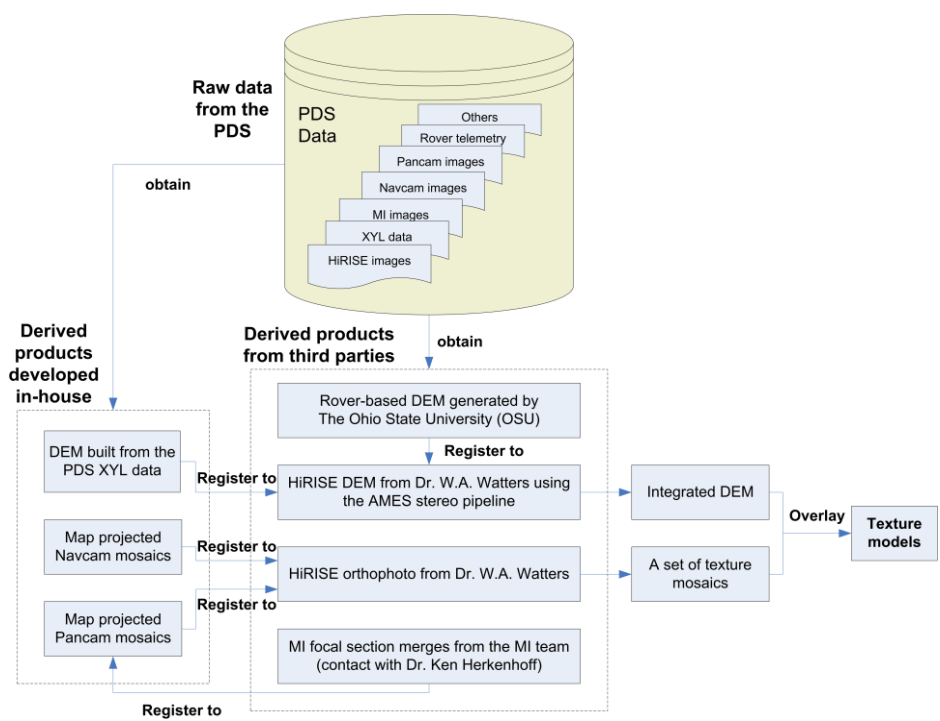

*4.1. Data Source*

A 2D image mosaic and 3D DEM (Digital Elevation Model) are two basic elements to build a texture model for the VA. The image mosaic and DEM are generated using data from Mars orbital and landed missions. The image data include HiRISE images from NASA MRO mission and the surface imagery taken by the Opportunity Rover.

An orbital DEM together with an orthophoto were created from HiRISE stereo images (image IDs: ESP_012820_1780, ESP_020758_1780) using the AMES stereo pipeline by Wes A. Watters (personal contact). The HiRISE orthophoto serves as the regional context. Each pixel in the HiRISE image has a resolution of ~25 cm. However, even with the high resolution of HiRISE data, relevant details at the scale of the Opportunity Rover are hard to discern. Rover-based images taken by Opportunity and topography data derived from these images are indispensable in providing required detail.

The ground-based image mosaics were generated with data taken by the Pancam, Navcam, and MI onboard the rover. Pancam is a multispectral and stereoscopic imaging system with a focal length of 43 mm and 16° field of view (FOV) [35]. Navcam consists of a pair of grayscale stereo cameras with a focal length of 14.67 mm and 45° FOV [36]. Pancam and Navcam acquire panoramic stereo images of the scene for both science and navigation purposes. The rover also has a MI camera that simulates a geologist's hand lens to provide a detailed view at a microscopic scale. MI has a 20 mm focal length lens, providing a ±3 mm depth of field and a $31 \times 31$ mm FOV at a 63 mm- working distance from the lens to the object being imaged [37,38]. Because of the limited depth-of-field of the MI's optics, a stack of MI images is taken at different target distances. The best-focused elements from all images in the given stack are then combined into a single image to provide the best focus over the entire field-of-view [38]. This derived MI image is known as a focal section merged image, which was generated for the rock targets Luis De Torres and Ruiz Garcia by the MI team (personal contact with Ken Herkenhoff). Pancam and Navcam raw data were obtained from the PDS.

*4.2. DEM Integration*

Several DEMs generated from orbital and rover-based data are used in the Santa Maria version of the VA. The HiRISE orbital DEM has a spatial resolution of a few decimeters covering the study area around 300 m × 300 m. The HiRISE DEM provides good regional control of the crater, but it lacks the detailed terrain information at the rover scale. Rover-based DEMs providing additional details of a rock or surface target are of interest. One source of the higher resolution DEM was generated by The Ohio State University (OSU) using data from a wide-baseline imaging campaign at Santa Maria. Wide-baseline stereo images are acquired by imaging the same region with Pancam with the rover sitting at two spots separated by about 5 m. These data are more accurate at distances further from the rover than those can be derived from normal stereo images with the fixed baseline between the cameras on the rover mast. The other type of rover-based DEM is derived from normal baseline stereo images. These data are standard products archived by the PDS [39]. Transforming these three sources of DEMs into the same reference frame and overlaying them together showed there was a tilt variation between the DEMs, especially in the east and north direction. This tilt was due to the error accumulated over time in the rover's position and orientation.

To remove the tilt variations among the DEMs, a set of operations were carried out to geo-reference the rover-based DEMs to the orbital derived data. These DEMs were first transformed to the same coordinate system. Then, a number of features found on both HiRISE DEM and Pancam wide baseline stereo DEM were manually selected. These features were served as control points for the data transformation. A rigid 3D transformation could not completely remove the tilt variation between orbital and ground DEMs. Rubber sheeting had to be applied to minimize the inconsistency between the data sets and to make smooth transition between the overlapping areas of both data sets. The two transformed data sets were integrated into a single DEM. After that, DEM obtained from the normal baseline stereo images was transformed in a similar manner to match the HiRISE and wide baseline data. In order to minimize the size of the terrain model, only subsets of normal baseline data covering the two rock targets were registered. Because the ground and orbital DEMs had different resolutions, a grid format for the merged DEM was large in size and limited the performance of the 3D visualization. A TIN (Triangulated Irregular Network) model was generated for the final fused DEM as shown in Figure 4.

**Figure 4.** Digital Elevation Model (DEM) integrated using rover-based and orbital data at Santa Maria Crater.

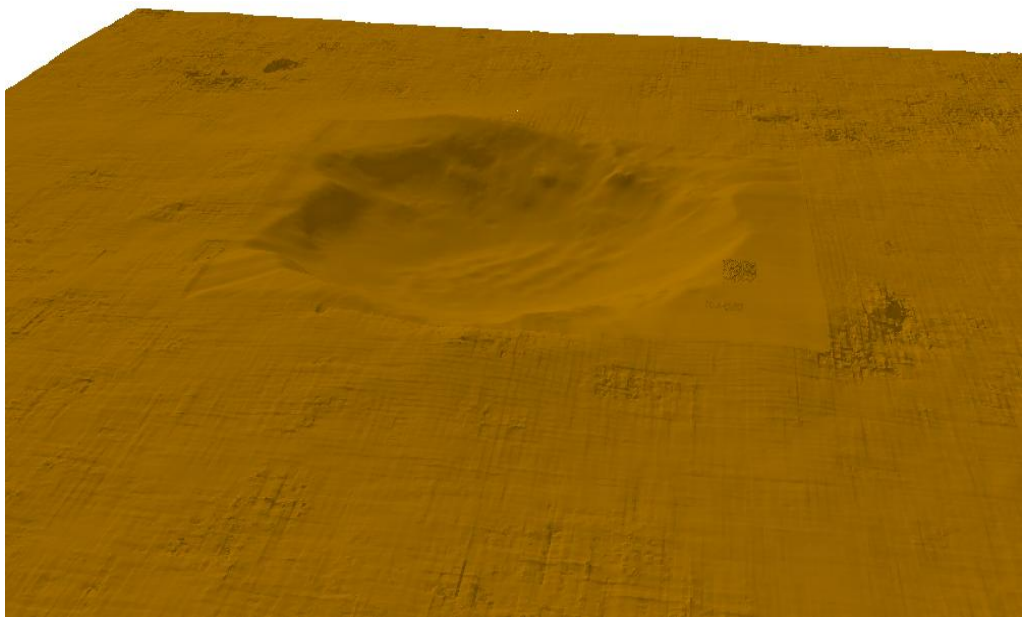

*4.3. Multi-Spectral Pancam Data Integration*

Rover based Pancam multi-spectral data are taken by 13 geology filters at wavelengths ranging from 400 to 1100 nm [35]. Six filters (L2 through L7), which are located on the left Pancam camera, have wavelengths in the visible range from blue to green to red. The other filters (R3 through R7) on the right camera cover in the near-infrared. Two additional filters on the right camera have redundant wavelengths in blue and red. The blue and red filters found on both cameras are used when acquiring stereo image pairs. These multispectral observations provide characterization of the mineralogy and color properties of the scene around the rover. The combination of images taken by the six visible wavelength filters can be used to construct an approximate true color picture of Mars [35]. Therefore, Pancam data provide the best source for the generation of a high-resolution texture mosaic.

Because of the limitation of rover resources during the mission, not every image sequence had Pancam data taken with all six visible wavelength filters. Images taken through filters L2 (753 nm), L5 (535 nm), and L7 (432 nm) or through filters L4 (601 nm), L5 (535 nm), and L6 (482 nm) were often selected to generate false color images. The six filters L2 through L7 were only used to generate approximate true color when available. In other cases the available filters were combined and assigned to the red, green, and blue channels to create approximate true color images or false color images.

A 360° Pancam panoramic view of the local terrain can consist of nearly 100 images in each filter. The images in these mosaics are often taken on different sols (Martian day) under slightly different lighting conditions. A MATLAB pipeline was built to create false color Pancam images and map them to the 3D terrain to generate textures models. The individual projected color images were mosaicked after contrast adjustment. Figure 5 shows one of the Pancam false color mosaics generated using data taken at the western rim of Santa Maria Crater. The color was manually adjusted between adjacent overlapping images in order to make a smooth transition of colors.

**Figure 5.** Pancam false color mosaic shows the interior of Santa Maria Crater. The floor of the crater is covered by sand ripples.

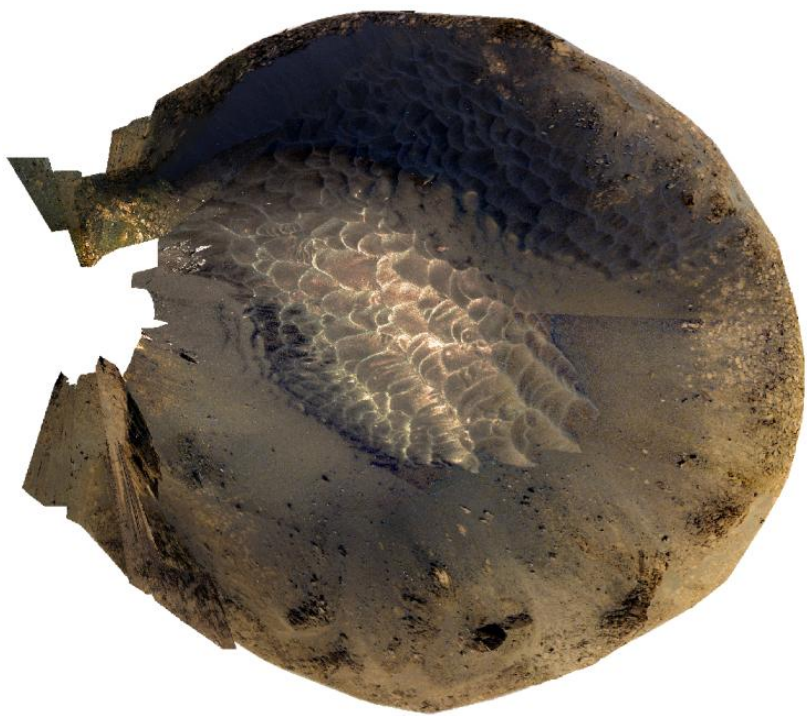

Additionally, Pancam multi-spectral data acquired from multiple camera positions were also combined to generate a big mosaic covering the study area, as shown in the color part of Figure 6a.

### 4.4. Merging Images from Multi-Instruments

Navcam image mosaics acquired by Opportunity provided nearly full coverage of the Santa Maria study area. Navcam images also have 3 times less resolution than Pancam data. The Pancam coverage was focused on the crater interior and the locations around the two rock targets, Luis De Torres and Ruiz Garcia. In order to generate a mosaic of the entire study area with the best resolution and color where available, the higher-resolution Pancam color image mosaics were resampled to match the resolution of Navcam mosaics. In addition, the tiny MI focal section merged images of the two rock targets were registered to Pancam images covering the rocks to provide context of the MI coverage relative to the surrounding parts of the rocks. HiRISE orthoimage had a complete coverage of the crater and was selected as the base control. The rover-based data were georeferenced to the HiRISE orthoimage by a number of registrations. The images were taken under different lighting conditions, so the contrast was adjusted slightly among images to make the color look consistent across image seams. Figure 6 shows the result of all the image mosaics registered in the same coordinate frame.

**Figure 6.** Image mosaics generated at Santa Maria Crater for the Virtual Astronaut. (**a**) A mosaic generated using HiRISE orthoimage, Pancam mosaics and Navcam mosaics provides the textural information for the VA; (**b**) A close up view of Luis De Torres shows a MI focal section merge image overlaid on a false color Pancam image. The area shown corresponds to the red box shown in (**a**).

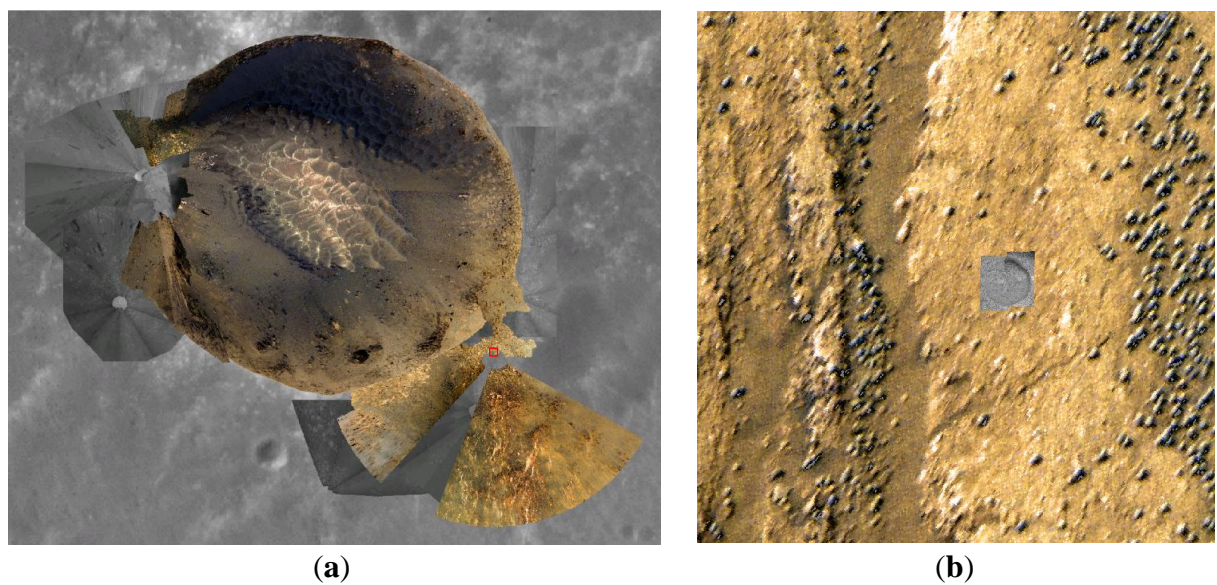

(**a**)                                                                          (**b**)

This mosaic combined with the integrated DEM discussed in section 4.2 was used for visualization in the 3D virtual environment. Figure 6b displays a MI focal section merge image taken at target Luis De Torres merged with a Pancam color mosaic. The circle in the MI image is a 5 cm × 5 cm hole ground into the rock by the Rock Abrasion Tool (RAT) mounted on the rover's robotic arm. Two views of the Santa Maria Crater area derived from Mars in Google Earth are shown in Figure 7 for comparison with the detail gained by combining orbital and ground-based images (Figure 6).

**Figure 7.** HiRISE textures shown in Mars in Google Earth. (**a**) A full-resolution HiRISE texture (~0.3 m/pixel) at Santa Maria Crater; (**b**) A zoom-in view of the textures at Luis De Torres and Ruiz Garcia.

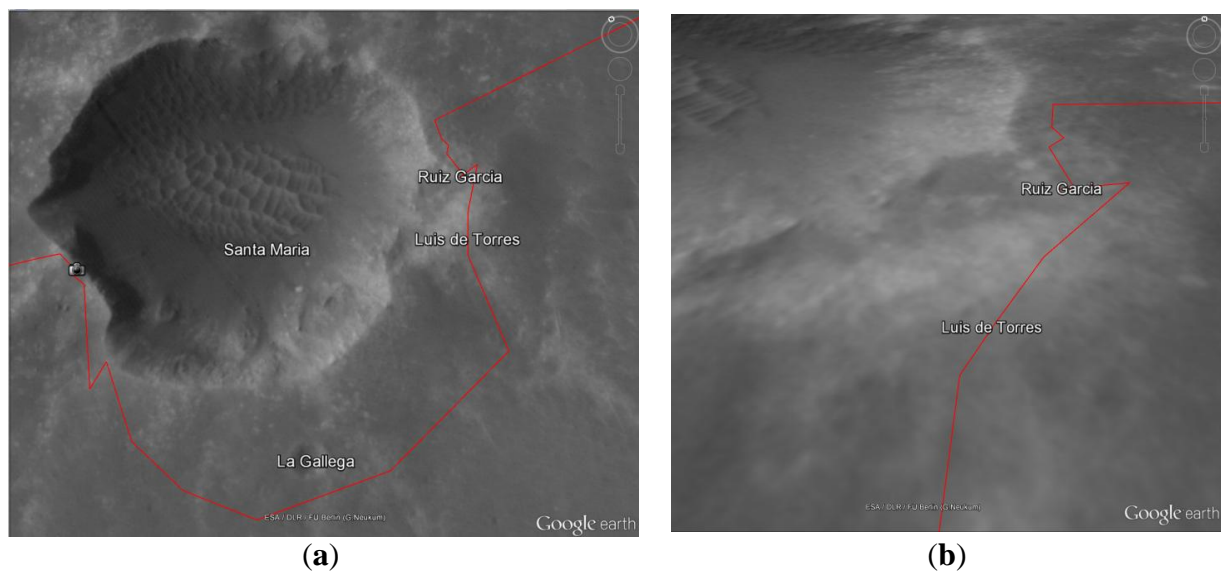

(**a**)                                                                                 (**b**)

## 5. Functionality of the Santa Maria VA

The Santa Maria VA supports three modes: walk, drive and target. Each mode gives the user a unique viewpoint to investigate the Martian surface. The Walk Mode allows the user to simulate freely along moving any direction around the terrain and to look left, right, up and down. The viewpoint is set at a few meters above the terrain. The viewpoint altitude is adjustable by the user. The Drive Mode simulates a rover driving along the actual traverse taken by Opportunity. The Target Mode lets a rover directly jump to one of the two rock targets with *in-situ* observations. Figures 8–11 are screen shots that illustrate the main functions of the Santa Maria VA. Figure 8 presents a close up view of the rover and the crater after the user used the Drive Mode and positioned the rover to the west of the crater. Figure 9 presents the scene as viewed by a user in Walk Mode entering into Santa Maria Crater at a height of 2.2 m above the terrain surface. The user can walk around the surface with mouse click. The screen's movement follows the mouse as the user points in the direction of desired movement.

**Figure 8.** A close up view of Santa Maria Crater with the rover model located at one of the positions that image data were acquired. The viewpoint is about 4.2 m above the terrain. The icons shown on top of the screen provide functional tools and controls. The yellow labels on the top-left of the screen give the position and orientation of the current viewpoint.

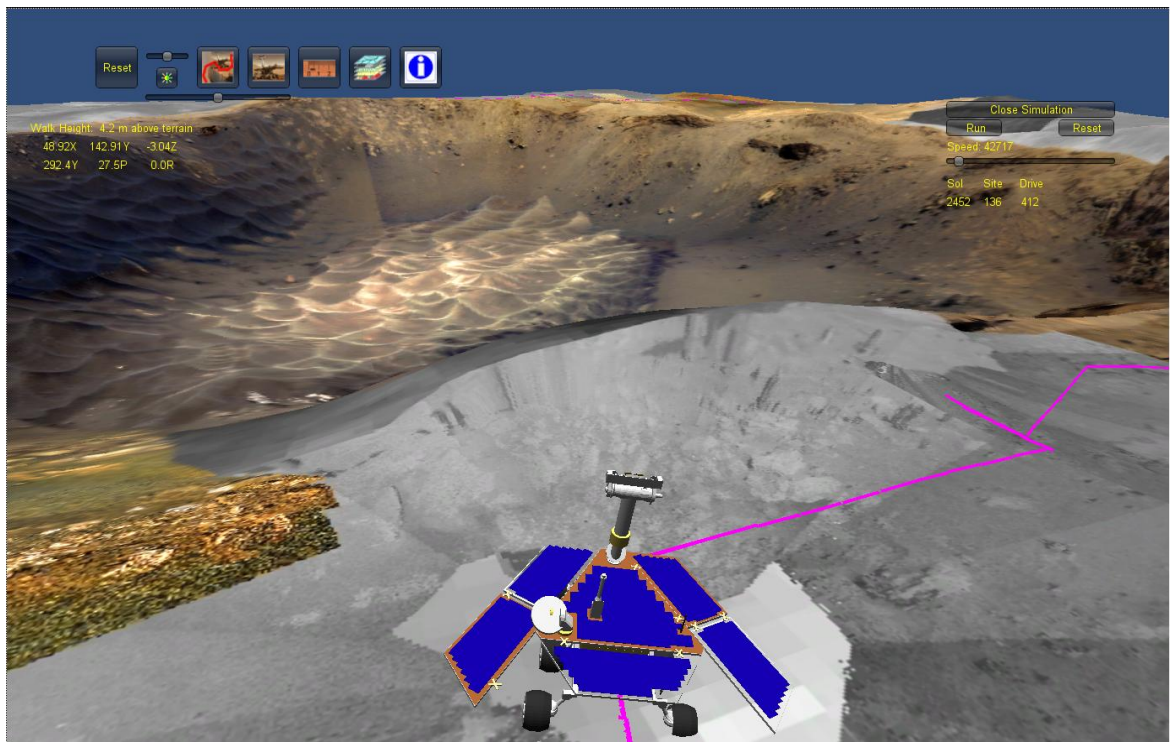

**Figure 9.** The interior of Santa Maria Crater is shown as an example of using the Walk Mode to simulate walking into the crater. The viewpoint is 2.2 m above the terrain surface.

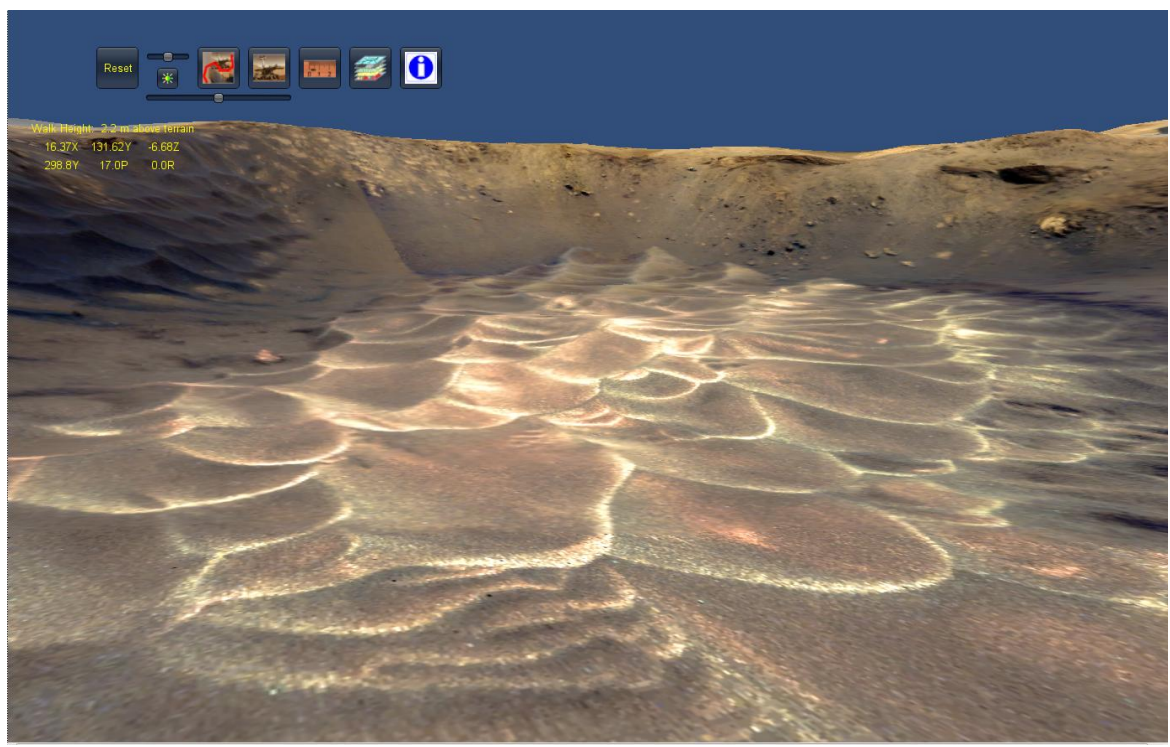

Target Mode in the VA allows a user to view data acquired on a surface target by the instruments that make *in-situ* measurements, such as the MI or APXS instruments. For the Santa Maria case, the two targets were Luis De Torres and Ruiz Garcia. Figure 10 is the observation of Rock Ruiz Garcia at Santa Maria Crater.

> **Figure 10.** In this example the rover is positioned where it was while making a series of *in-situ* measurements. The menu on the right side of the screen expands to include a list with the MI focal section merge and Hazcam images available for viewing. Selecting one of the menu entries will display the image in a pop-up window. In this example, the pop-up window shows the image taken by the rover front Hazcam camera to document the placement of the MI instrument on the target. A red arrow shown above the rock is pointing to the target where the MI experiment was carried out. Functions are provided in the pop-up window to open the full-resolution image or link to the Mars Exploration Rover (MER) Analyst's Notebook for data downloading.

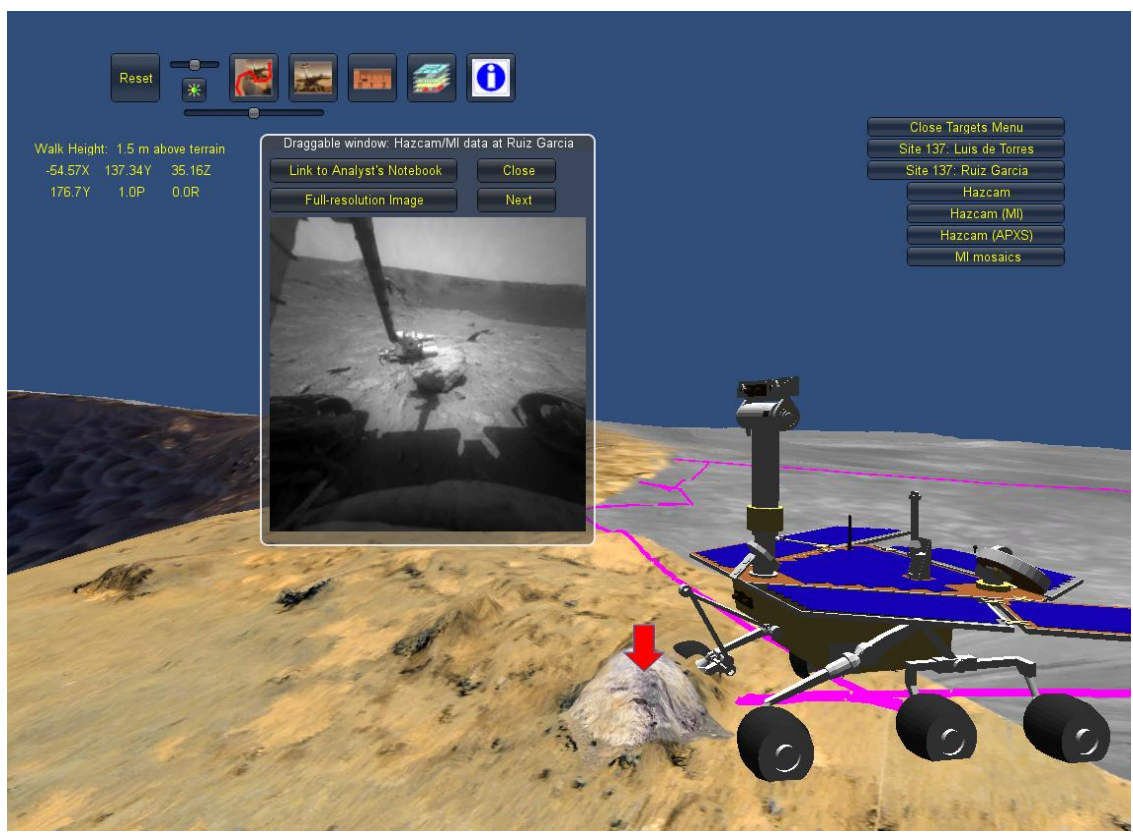

A fourth function of the Santa Maria VA is the ability to measure the distance between two points. This allows the user to measure the size of objects like rocks in the scene. To use this function the user first left click on a point in the scene and then drag the mouse to a second point. Both points are marked in red on the screen for reference. The system will then calculate the distance between two points and display the result on the screen in a pop-up window as shown in Figure 11.

**Figure 11.** A user makes a measurement of distance between the two red points on the 3D scene. Results are shown in a pop-up window attached to the last position of the mouse. The distance between the points in this example is about half a meter, and the height difference is about 20 cm. Both points are about 2.5 m away from the viewer.

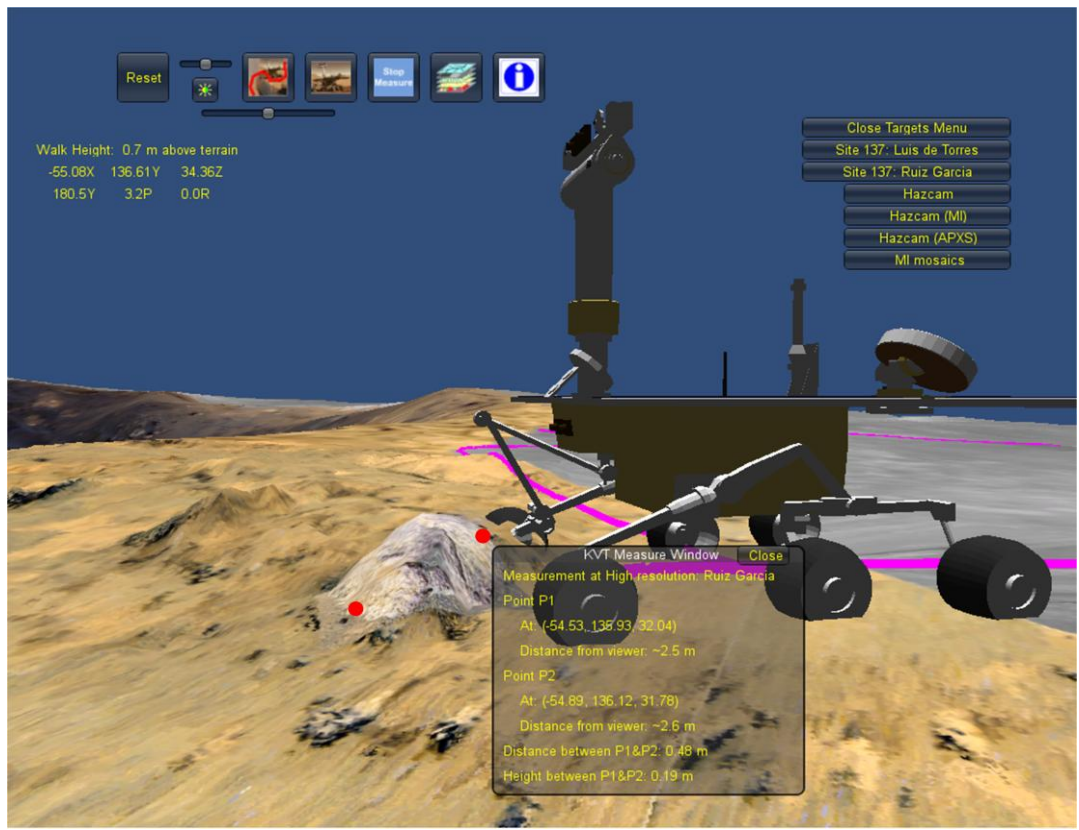

## 6. Conclusions

The initial version the Virtual Astronaut created for Santa Maria Crater [40] demonstrates that the concept can provide an effective tool for organizing and presenting data from multiple instruments that have differing resolutions in an interactive 3D virtual environment. The VA integrates data from both orbital and landed missions. The VA allows the user to explore and view the surface at different scales. The user can further interact with the environment by viewing and accessing *in-situ* observations, making measurements of features, and controlling an animation of rover drives. Within the VA tool, the user can freely move around the scene and look in any direction desired. Thus, the user is not limited to seeing the scene only through the viewpoint of the rover. The high-resolution image mosaics built from rover-based combined with orbital mosaics for regional coverage makes the VA unique when compared with other Mars virtual reality systems. The development tools chosen to implement the VA provided an efficient environment for iteratively creating functionality and revising these functions based on feedback from users. Future work includes generating additional versions of the VA as Opportunity explores the rim of Endeavour Crater and Curiosity explores Gale Crater.

## Acknowledgments

This research has been carried out through funding provided by the NASA's PDS Geosciences Node. The authors would like to thank all the colleagues working at the Geosciences Node for valuable comments. Especially comments from Raymond E. Arvidson are appreciated. We would like to thank the Geosciences Node Advisory Group for their valuable feedback and comments. Special thanks are given to OSU Mapping and GIS Laboratory, Wes A. Watters from Cornell University for the valuable data they provided for Santa Maria Crater.